\documentclass[conference]{IEEEtran}
\ifCLASSINFOpdf
\usepackage[pdftex]{graphicx}
\else
\fi
\usepackage[cmex10]{amsmath}
\usepackage{amsfonts}

\usepackage{array}
\begin{document}
%
\title{Refining Adverse Drug Reactions using Association Rule Mining for Electronic Healthcare Data}


\author{\IEEEauthorblockN{Jenna M. Reps, Uwe Aickelin}
\IEEEauthorblockA{School of Computer Science\\
University of Nottingham\\
Email: jenna.reps@nottingham.ac.uk\\
Email: uwe.aickelin@nottingham.ac.uk}
\and
\IEEEauthorblockN{Jiangang  Ma}
\IEEEauthorblockA{College of Engineering \& Science\\
Victoria University\\
Melbourne\\
Email: Jiangang.Ma@vu.edu.au}
\and
\IEEEauthorblockN{Yanchun Zhang}
\IEEEauthorblockA{School of Computer Science\\
Fudan University\\
Email: yanchunzhang@fudan.edu.cn}
}


%


\maketitle

\begin{abstract}
Side effects of prescribed medications are a common occurrence. Electronic healthcare databases present the opportunity to identify new side effects efficiently but currently the methods are limited due to confounding (i.e. when an association between two variables is identified due to them both being associated to a third variable).

In this paper we propose a proof of concept method that learns common associations and uses this knowledge to automatically refine side effect signals (i.e. exposure-outcome associations) by removing instances of the exposure-outcome associations that are caused by confounding. This leaves the signal instances that are most likely to correspond to true side effect occurrences. We then calculate a novel measure termed the confounding-adjusted risk value, a more accurate absolute risk value of a patient experiencing the outcome within 60 days of the exposure.

Tentative results suggest that the method works. For the four signals (i.e. exposure-outcome associations) investigated we are able to correctly filter the majority of exposure-outcome instances that were unlikely to correspond to true side effects. The method is likely to improve when tuning the association rule mining parameters for specific health outcomes.

This paper shows that it may be possible to filter signals at a patient level based on association rules learned from considering patients' medical histories. However, additional work is required to develop a way to automate the tuning of the method's parameters.

\end{abstract}


%
\IEEEpeerreviewmaketitle

\section{Introduction}
Medications are prescribed to patients suffering from morbidities with the aim of improving the health of patients health. Unfortunately, the majority of medications will also induce some negative side effects referred to as adverse drug reactions (ADRs). Clinical trials and other research present a clear view of the positive effects of medication but are often insufficient for obtaining information about ADRs \cite{Amery1999}. This has prompted researchers to develop methods that can identify unknown ADRs efficiently via a process known as pharmacovigilance. 

The first stage of identifying an ADR is to generate a signal, a previously unknown association between an exposure (e.g. a drug) and health outcome. If an exposure causes an outcome, then the exposure and outcome will be associated. However, only because an exposure and an outcome are associated, they may not have a causal relationship, e.g. there could be a third confounding factor. As a consequence, it is important to filter ADR signals (associations) and remove those that are unlikely to correspond to causal relationships (true ADR will be a causal relationship). Therefore, the generated signals need to be actively monitored during the process of signal refinement. The signals that are most likely to correspond to ADRs or where the corresponding health outcomes are very severe (e.g., death) are then evaluated via formal epidemiological investigations.

Signal generation using electronic healthcare databases has been a recent focus of attention and various methods have been presented \cite{Reps2014,Jin2006,Noren2010}. Due to the high false positive rates of these signal generation methods (i.e. they often signal drug and health outcome associations that are non-causal) \cite{Ryan2012,Reps2013}, it is imperative that new methods are developed that can further filter the signals and reduce the false positive rate. This would enable the signal evaluation to occur at an earlier point in time and decrease the overall time it takes to come to a conclusion about the signal.

Recently, a refinement method using a logistic regression has been proposed \cite{Caster2013}. This method fits a logistic regression to predict the occurrence of the health outcome using a binary independent variable indicating whether the drug that is potentially responsible for causing the health outcome is present and then fits another similar logistic regression but incorporates additional independent variables that may be confounding the relationship between the drug and health outcome. If the weight given to the drug indication variable in the trained logistic regression decreases when potential confounders are considered, then the authors argue this suggests that the drug is unlikely to cause the health outcome. Justification is that if the drug causes the health outcome then the parameter representing the increase in odds ratio caused by taking the drug should be consistent and not be affected by including additional covariates into the model. Although this method has merit, the downsides are that it is difficult to automate. It requires identifying possible confounders and training multiple logistic regressions for each signal. This is time consuming. 

Other researchers have argued that the majority of health outcomes suspected of being ADRs are due to pre-existing illnesses. To remove these suspected ADRs that are caused by pre-existing illnesses, natural language processing techniques have been applied to written reports often found within electronic healthcare databases \cite{Haerian2012}. It was shown that these reports can be used to reduce the false positive rates, but the written reports may not always be available.  Alternative methods need to be developed for when written reports are absent.

Our starting point is to consider more closely how a medical expert works and if we can automate this. Experts are able to apply their knowledge and experience to identify when a patient is the likely recipient of an ADR. They can look at a patient's medical history and current state to identify possible alternative causes of the health outcome. This prompts the idea of using association rule mining to identify common patterns involving drugs and health outcomes and then filter instances when a signal's corresponding health outcome has a potential alternative explanation. This novel association rule refinement method does not require written reports, although it would probably only be suitable for electronic healthcare databases with a few years of patient history. 

The objective of this paper is to present a proof of concept detailing how association rules \cite{Agrawal1993} could be used to automatically refine ADR signals. In addition to the proof of concept, tentative results are presented for the automatic refinement when considering four signals that have occurred within The Health Improvement Network database for the quinolone drug family. 

\section{Materials and Methods}
\subsection{The Health Improvement Network}
The health improvement network (THIN) database is a UK general practice database containing medical and prescription records for millions of patients. The validity of the database has been evaluated \cite{THIN} and it seems a suitable source for pharmacovigilance. The records have a date stamp, but the time that each record was entered is not recorded. The THIN database also contains information about the patients including their year or births, their genders and summary details of the demographics within their practices. We decided to partition the THIN database into two sets. One set contains complete medical records for a subset of randomly selected patients that will be used to develop novel methods and the other set contains complete medical records for the remaining patients that will be used to evaluate and validate the new methods that are developed. Therefore, in this paper we use the first subset of patients within the THIN database and this corresponds to a total of approximately 4 millions patients. The THIN database contains two hierarchical structures. Health outcomes are recorded via a tree structure known as Read codes and prescriptions have an associated British National Formula (BNF) code \cite{BNF} that also has a tree structure. 

\begin{figure}[t]
\centering 
\includegraphics[trim=0cm 0cm 0cm 0cm, clip=true, width=0.5\textwidth]{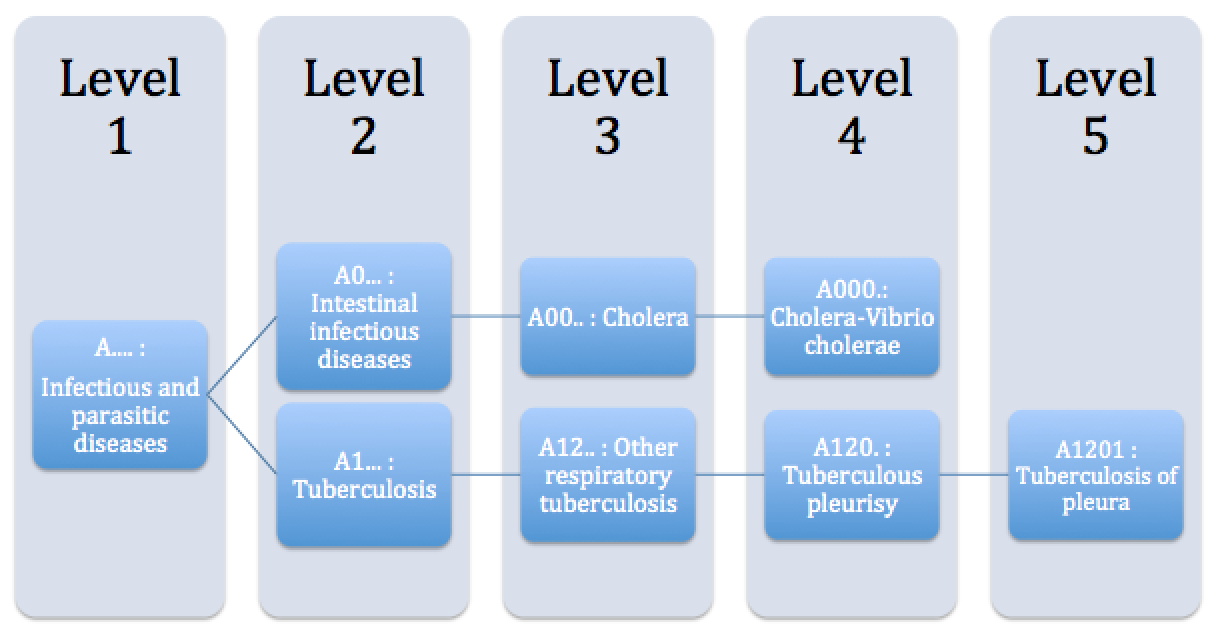}
\caption{An overview of the read code hierarchy.} 
\label{read_code_pic}
\end{figure}

The Read codes are used to recorded health outcomes and have five levels of detail. Each Read code corresponds to a health outcome description such as an observation, a procedure or an administrative event. The Read code consists of five values, $x_{1}x_{2}x_{3}x_{4}x_{5}$, where each value is either from the upper-case Latin alphabet, from the lower-case Latin alphabet or a dot, $x_{i}\in \{A-Z,a-z,\bullet\}$. The level of a Read code $X$ is defined as,
\begin{equation}
l(X)=\arg\max_{i} \{i \in [1,5] | x_{i} \neq \bullet\}
\end{equation}
The higher the Read code level, the more specific the corresponding medical event description is. The Read codes have a tree structure. For Read code $X$, its direct parent's Read code $Y$ is,
\begin{equation}
y_{i}= \left\{ \begin{array}{ll} x_{i} & \quad \mbox{if }i< l(X) \\
\bullet & \quad \mbox{otherwise} \end{array} \right.
\end{equation}
and its level $k \in [1,5]$ Read code, denoted $\hat{X}|_{k}$, is,
\begin{equation}
\hat{x}_{i}= \left\{ \begin{array}{ll} x_{i} & \quad \mbox{if }i\leq k \\
\bullet & \quad \mbox{otherwise} \end{array} \right.
\end{equation}
For example, considering the Read code $A11zz$, its direct parent is $A11z\bullet$, its level 3 parent is $A11\bullet\bullet$, its level 2 parent is $A1\bullet\bullet\bullet$ and its level 1 parent is $A\bullet\bullet\bullet\bullet$. A child Read code corresponds to a more detailed health outcome than its parent's Read code. The Read code $A\bullet\bullet\bullet\bullet$ corresponds to `infectious and parasitic infections', whereas the child Read code $A1\bullet\bullet\bullet$ corresponds to ` Tuberculosis'. An examples is presented in Figure \ref{read_code_pic}.

The BNF codes also have a hierarchical tree structure. Each BNF code consists of four levels in the form, $b_{1}.b_{2}.b_{3}.b_{4}$, where each $b_{i} \in \mathbb{Z}$. The BNF codes show a relationship between drugs based on their indications (causes of taking the drug). The level $k \in [1,4]$ BNF code of $B$, denoted $\hat{B}|_{k}$, is calculated by,
\begin{equation}
\hat{b}_{i}= \left\{ \begin{array}{ll} b_{i} & \quad \mbox{if }i\leq k \\
0 & \quad \mbox{otherwise} \end{array} \right.
\end{equation}

Validation steps are implemented when the THIN data are being collected but issues are still known to arise. The main problem is due to patients changing general practice, e.g. by moving home. Patients are not tracked within the THIN database, so if a patient moves practice, he will be assigned a new unique ID. It is common for newly registered patients' pre-existing illnesses to be re-entered into THIN with 'incorrect' (i.e. new) dates. For example, after a patient registers at a new practice, his doctor is likely to enter his pre-existing illnesses when the doctor finds out about them during his initial visits, but the dates of these records will be different to the actual dates when the illnesses started. To prevent this biasing the analysis, the first 12 months of a newly registered patient's drug records are excluded from the analysis of identifying drug side effects. Previous studies showed that the bias is sufficiently reduced when the first 12 months of medical records are removed \cite{Lewis2006}. Drug prescriptions recorded within the last month of the THIN database being generated are also excluded from the analysis of identifying drug side effects to prevent under-reporting. Thi sis because if a patient had a drug recorded on the last day that the records were collected for the version of the THIN database used in the study, then it will not be possible to have a full record of any side effect that may have occurred.

\subsection{Association Rule Mining}
Association rules mining is frequently applied to transaction data, such as items purchased at a supermarket, to find rules of the form \{antecedent\} $=>$ \{consequence\} (read as a set of antecedent events that frequently precede the occurrence of a set of consequent events) that occur in shopping baskets. An example of an association rule is, 
\{pasta, tomato sauce\} $=>$ \{cheese\}, where this rule can be considered to mean shoppers that buy pasta and tomato sauce are also likely to buy cheese. In a similar manner, patients’ medical baskets can be constructed based on the records they have in the THIN database and association rule mining can be applied to find medical association rules. Due to the number of possible rules being very large, association rule are often restricted so that only interesting rules are discovered. 

The formal problem of association rule mining is: Let $I=\{ i_{k}, k \in [1,n] \}$ be a set of $n$ items and let $D=\{ t_{k} \subset I, k \in [1,m] \}$ be a set of $m$ baskets called the $database$. An association rule is an implication, $X=> Y$, where $X,Y \subset I$ are itemsets and $X \cap Y = \emptyset$. $X$ is referred to as the antecedent and $Y$ is the consequence.

In shopping data, interesting rules are generally those that will be found in lots of shopping baskets and the probability of a basket containing the consequent event given the antecedent is high. To mine these sorts of rules two common measures are used, the support and the confidence. 

{\bf Support:} The support of an itemset $X$ is the proportion of baskets within the database that contain $X$,
\begin{equation}
supp(X) = |\{t_{i}\in D | X \subset t_{i} \}|/m,
\end{equation}
where the cardinality of $X$ is denoted $|X|$ (the number of items in the itemset). The support of an association rule $X=> Y$ is,
\begin{equation}
supp(X=> Y) = |\{t_{i}\in D | X \cup Y \subset t_{i} \}|/m
\end{equation}
this is approximately the probability of finding both the itemset $X$ and itemset $Y$ within a randomly chosen basket.

{\bf Confidence:} The confidence of a rule $X => Y$ is the proportion of baskets that contain the antecedent and consequence, $X \cup Y$, divided by the proportion of basket that contain the antecedent $X$,
\begin{equation}
conf(X => Y) = supp(X \cup Y)/supp(X)
\end{equation}
The confidence of an association rule $X=> Y$ is approximately the conditional probability of a randomly chosen basket containing the itemset $Y$ given it contains the itemset $X$.

The support and confidence are used as follows. The user inputs a minimum support constraint and a minimum confidence constraint that restricts the output rules to those that have a support greater than the minimum support and a confidence greater than the minimum confidence. This is not suitable for mining medical baskets as we are still interested in rules that occur relatively less often, so the minimum support constraint may prevent many rules of interest being found. For example, if a health outcome is relatively rare, then the support of medical itemsets that contain the health outcome are unlikely to be greater than a minimum support unless is it set very low (which then causes computational efficiency issues). 

{\bf Left Support:} When a consequence item occurs rarely in the database, rather than using the support constraint the left support constraint is often more suitable. The left support is the number of shopping baskets that contain the antecedent divided by the number of shopping baskets,
\begin{equation} 
leftSupp(X => Y) = supp(X)
\end{equation}
The left support of the association rule $X=> Y$ is the same as the support of the itemset $X$. By using the left support constraint, it is possible to mine association rules $X=> Y$ where the itemset $Y$ is rarely contained in a basket. For example, if the itemset $X$ has a support of 0.01, the itemset $Y$ has a support of 0.001 and $X$ is always found in a basket that contains $Y$. The rule $Y=> X$ is interesting and it could be used to help predict $X$, but it only has a support of 0.001. If a minimum support of 0.005 was used to find the association rules then this rule would not be found. However, if the minimum left support constraint of 0.005 was used, then the rule would be identified (as the left support is 0.01). This shows the advantage of using the left support when a consequence is rare (i.e. has a low support). 

{\bf Lift \& Chi-squared:} Two more measures of interestingness that may be useful for detecting medical association rules are the lift (a measure of association) and the chi-squared value (the significance of the association),
\begin{equation}
lift(X => Y) = supp(X \cup Y)/(supp(X) \times supp(Y))
\end{equation}
\begin{equation}
\chi^2= \sum (O_{i}-E_{i})^2 /E_{i}
\end{equation}
The lift calculates the number of baskets that the association rule was found in divided by the number of baskets you would expect the rule to be found in if the itemsets $X$ and $Y$ were independent. A large lift indicates that there is a dependency between itemsets $X$ and $Y$. The chi-squared value of an association rule, $X=> Y$, uses the observed $O_{i}$ and expected $E_{i}$ values as described in contingency tables \ref{obs}-\ref{exp},
\begin{table} \centering
\caption{The observed contingency table for the association rules $X=> Y$.}
\label{obs}
\begin{tabular}{c|cp{0.2\textwidth}}
& Y & not Y \\ \hline
X & O$_{1}$ & O$_{2}$ \\
& ($supp(X\cup Y)$) & ($supp(X)-supp(X\cup Y)$) \\
not X & O$_{3}$ & O$_{4}$ \\
& ($supp(Y)-supp(X\cup Y)$) & ($1-supp(Y)-supp(X)+ supp(X\cup Y)$ ) \\
\end{tabular}
\end{table}
\begin{table} \centering
\caption{The expected contingency table for the association rules $X=> Y$.}
\label{exp}
\begin{tabular}{c|cp{0.2\textwidth}}
& Y & not Y \\ \hline
X & E$_{1}$ & E$_{2}$ \\
& ($supp(X)supp(Y)$ ) & ($supp(X)(1-supp(Y))$) \\
not X & E$_{3}$ & E$_{4}$ \\
& ($supp(Y)(1-supp(X))$) & ($1-supp(Y)-supp(X)+ supp(X)supp(Y)$) \\
\end{tabular}
\end{table}

To generate association rules for the THIN database we consider the following items: gender, Read codes up to level 3 (i.e. a level 4 or level 5 Read code of the form $x_{1}x_{2}x_{3}x_{4}\bullet$ and $x_{1}x_{2}x_{3}x_{4}x_{5}$, respectively, is mapped to its parent level 3 Read code $x_{1}x_{2}x_{3}\bullet\bullet$) and BNF codes up to level 2 (i.e. any level 3 or 4 BNF code of the form $y_{1}.y_{2}.y_{3}.0$ or $y_{1}.y_{2}.y_{3}.y_{4}$ is mapped to its parent level 2 BNF $y_{1}.y_{2}.0.0$ ). This mapping was done to reduce the number of items and speed up the association rule discovery. A medical basket is created for each patient in the THIN database who was active for a total of 24 months or more containing their gender, all their level 3 mapped Read codes and level 2 mapped BNF codes recorded in the database. The association rule mining is then applied to the $database$ containing these medical baskets with a minimum left support constraint of $0.001$ and a minimum confidence support of $0.01$. The set of all association rules mined from the $database$ satisfying the constraints is denoted by $THINarules$. These parameter values are based on discussions with clinical experts.

The minimum confidence of $0.01$ means only rules where a minimum of 1\% of the baskets containing the antecedent also contain the consequence are returned. The value was chosen to be small, as we want to investigate the confidence values for the rules found in this paper. A consequence that occurs in less than 1\% of people with the antecedent is not useful as it is rather unlikely that a patient will experience the consequence given they have the antecedent, so the rule is not useful for prediction the consequence. Using a minimum left support of $0.001$ means we only get rules where the antecedent occurs for 1 in 1000 patients, this was chosen due to the minimum confidence. As there are 4 million baskets, the rules of interest require that 4000 baskets contain the antecedent when the minimum left support is $0.001$. A minimum of 40 of these baskets must also contain the consequence when the minimum confidence is $0.01$. Therefore using these constraints ensure that the rule occurs in a minimum of 40 baskets. If the minimum left support was reduced to 1 in 10000, then we could get rules that were only found in 4 baskets. These rules are likely to occur by chance and the majority of the rules mined will be useless. Therefore we decided the minimum support of 0.001 was suitable and this also ensured the useful association rules were mined efficiently. 

\subsection{Novel Refinement Method}
\begin{figure}[t]
\centering 
\caption{An overview of the refinement framework that generates the confounding-adjusted risk for any DOI-HOI selected.} 
\label{meth}
\includegraphics[trim=0cm 8cm 0cm 0cm, clip=true, width=0.5\textwidth]{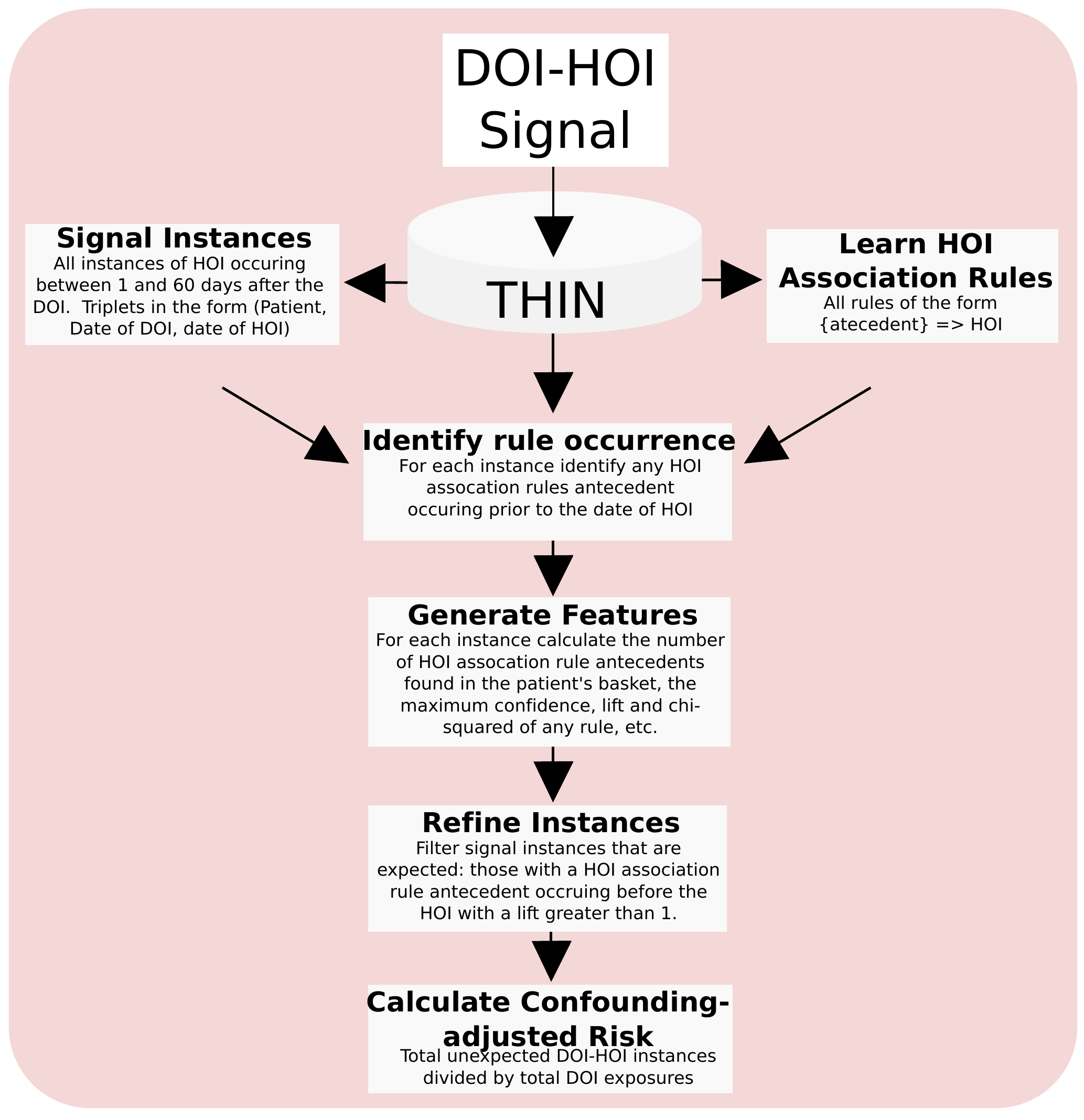}
\end{figure}

The proposed refinement method is an unsupervised technique that takes as input an ADR signal consisting of a drug of interest (DOI) and health outcome of interest (HOI) pair (DOI-HOI). The method then returns a refinement score for the DOI-HOI pair corresponding to a confounded-adjusted risk value. This score can be used to identify which ADR signals should be formally evaluated by epidemiological hypothesis testing methods. The refinement method finds all the instances in the database where a patient is prescribed the DOI and then experiences the HOI within two months. In summary, the method works by filtering any instance of the DOI-HOI where the HOI is expected based on other medical events that the patient experienced prior to the HOI. The absolute risk of the HOI occurring within two months of the DOI exposure is then calculated, but the occurrences of any HOI that is expected are ignored. This then leaves us with the confounding-adjusted risk value. Removing 'expected' HOIs was previously considered in the MUTARA method \cite{Jin2006}, but they only consider a HOI to be expected when a patient had the same HOI during some time period shortly before the drug. They do not consider determining expectation based on using associations with other drugs, health outcomes or based on gender. An overview of the novel association rule based methodology is presented in Figure \ref{meth}.

The step-by-step process for the refinement is:
\subsubsection{Identify signal instances}
The instances of a DOI-HOI signal are the cases when a patient is prescribed the DOI for the first time and experiences the HOI between 1 and 60 days after. These are the patients that appear to experience the potential ADR. The set of signal instances consists of triplets containing the patient experiencing the DOI-HOI signal, the date that the DOI was recorded for the patient and the date that the HOI was record for the patient such that the date of the HOI lies between 1 and 60 days after the date of the DOI. 

In this paper the DOI is the quinolone drug family. We selected four different HOIs that were associated with quinolones when considering the after/before (AB) ratio. The AB ratio is a basic measure of association that calculates the number of distinct prescriptions of the quinolones that have the HOI recorded during 1 and 60 days after the prescription divided by the number of distinct prescriptions of the quinolones that have the HOI recorded during 60 and 1 days before the prescription. This measure of association is quick to calculate but generates numerous signals that require refining. Table \ref{sig} presents the number of instances for each DOI-HOI signal investigated in this paper.

\begin{table*} \centering \caption{The quinolone-HOI signals investigate in this paper.} \label{sig}
\begin{tabular}{ccccc}
HOI &Read code & AB ratio & Instances & HOI rules \\ \hline \hline
Candidiasis & AB2 &1.27 &3388 &6251720 \\
Sudden death, cause unknown & R21& 73& 73& 11219 \\
Multiple organ failure & C19& 9& 9 & 1 \\
Secondary malignant neoplasm & B572& 5& 5& 216262 \\
\end{tabular}
\end{table*}

\subsubsection{Identify HOI association rule occurrence}
We extract the HOI association rules from the association rules mined from the THIN database. As the association rules contained in $THINarules$ only contain level 3 versions of a Read code, when refining a level 4 or 5 version we use their level 3 HOI's association rules. The HOI association rules are all the rules $\{antecedent\} => \{HOI\} \in THINarules$. These are the rules with the HOI as the consequence.

For each DOI-HOI instance, we then investigate the items (level 3 Read code, level 2 BNF and gender) recorded for the patient up to the date that the HOI occurred and identify if the patient had the items corresponding to any HOI association rule’s antecedent recorded. This tells us whether their recording of the HOI has a plausible non-DOI explanation. The measures such as confidence, chi-squared and lift of each identified HOI rule give insight into how likely the alternative causes are to blame for the patient's HOI occurrence rather than the DOI. 

In detail, for each instance we determine whether the patient's medical basket composed of the items recorded before the HOI contains any HOI association rule's antecedent, we calculate the maximum lift, the maximum confidence and the maximum chi-squared values when considering all the HOI association rules where all the antecedent items are recorded for the patient before the HOI was recorded. So for each instance of the DOI-HOI, we have a binary value indicating whether there was any alternative cause of the HOI, and three real values that given insight into the association between any alternative cause and the HOI. 
\subsubsection{Refine instances}
Using the features we consider the HOI as `expected' for any instance where an alternative potential cause of the HOI was identified and the alternative cause corresponded to an antecedent of a HOI association rule with a lift greater than 1. This lift was chosen as this indicates whether the antecedent is associated to the HOI. Therefore, any antecedent corresponding to a HOI association rule with a lift greater than 1 is likely to be a confounder for the HOI.

Overall summary values for the DOI-HOI are then calculated based on aggregating the DOI-HOI instance values. These include the number of instances that have an alternative cause, the number of instances that have a maximum lift greater than 1 and the average of the instances’ maximum confidence and chi-squared values. This is the first study aiming to use association rules to refine DOI-HOI instance, so we do not know which association rule measures will be most informative.  By calculating different measures in this study we can gain insight into the measures that are useful.

\subsubsection{Calculate confounding-adjusted risk}
The confounding-adjusted risk is calculated as the number of unexpected instances of the DOI-HOI divided by the number of instances of the DOI.

\subsection{Example}
The refinement methodology is now detailed using a made up example of a medical database. Assume we are investigating the signal DOI1-HOI5. We find that 25 patients are prescribed DOI1 within the database and we extract a subset of the database containing all the patient records for patients who have HOI5 recorded within 1 and 60 days from DOI1, this is presented in Table \ref{ex}.
\begin{table}
\centering
\caption{ This is a table of the records for the instances experiencing the signal DOI1-HOI5 within the made up example database.}
\label{ex}
\begin{tabular}{cccc}
RecordID & Patient & Item & Date \\ \hline \hline
1 & 1 & DOI1 & 05/06/2003 \\
2 & 1 & DOI2 & 06/07/2003 \\
3 & 1 & HOI1 & 07/07/2003 \\
4 &1 & DOI2 & 01/08/2005 \\
5 &1 & HOI5 & 01/08/2005 \\
6 & 1 & HOI5 & 08/09/2005 \\
7 & 2 & HOI3 & 15/01/1999 \\
8 & 2 & DOI2 & 17/01/1999 \\
9 & 2 & DOI1 & 28/06/2001 \\
10 & 2 & HOI5 & 14/08/2001 \\
11 & 2 & HOI3 & 27/01/2005\\
12 & 3 & DOI2 & 23/11/2009 \\
13 & 3 & DOI2 & 23/1/2010 \\
14 & 3 & HOI3 & 19/3/2010 \\
15 & 3 & DOI2 & 21/3/2010 \\
16 & 3 & DOI1 & 21/3/2010 \\
17 & 3 & HOI5 & 22/3/2010 \\
18 & 3 & HOI2 & 22/3/2010\\
19 & 4 & DOI1 & 1/1/2011 \\
20 & 4 & HOI5 & 5/1/2011 \\
\end{tabular}
\label{examp}
\end{table}

The DOI1-HOI5 instances are: (1,05/06/2003,01/08/2005), (2,28/06/2001,14/08/2001), (3, 21/3/2010,22/3/2010) and (4, 1/1/2011, 5/1/2011) as we only consider the first time a patient is prescribed DOI1. Using the whole THIN database we find association rules of the form \{antecedent\}$=>$ \{HOI5\}. The HOI5 rules extracted are presented in Table \ref{ex_rules}.
\begin{table}[t] 
\centering
\caption{Example HOI5 rules extracted.}
\label{ex_rules}
\begin{tabular}{ccccccc}
ruleID & Antecedent & Consequence & $leftSupp$ & $conf$ & $\chi^{2}$ & $lift$ \\ \hline \hline
1 & DOI2 & HOI5 & 0.0012 & 0.03 & 200 & 1.4 \\
2 & HOI27,DOI2 & HOI5& 0.001 & 0.02 & 150 & 1.1 \\
3 & HOI3& HOI5 & 0.0015 & 0.013 & 160 & 1.5 \\
4 & HOI3, DOI2& HOI5 & 0.001 & 0.02 & 134 & 1.3\\
\end{tabular}
\end{table}

For each instance, we now search their items up to the point the record of HOI5 suspected to being caused by DOI1 is recorded and identify whether any of the HOI5 rule antecedents are present. For the instance (1, 05/06/2003, 01/08/2005), the HOI5 antecedents of ruleID 1 are present, so for this instance there is 1 HOI5 association rule, the maximum confidence, chi-squared and lift is therefore 0.03, 200 and 1.4 respectively. The records for instance (2, 28/06/2001, 14/08/2001 ) up to the HOI5 recorded on 14/08/2001 contain the antecedents of ruleIDs 1,3 and 4. Therefore the total number of HOI5 association rules is 3, the maximum confidence is 0.03 (the max out of 0.03 ,0.013 and 0.02), the maximum chi-squared is 200 (max out of 200, 160, 134) and the maximum lift is 1.5 (max out of 1.4, 1.5, 1.3). The instance (3, 21/3/2010, 22/3/2010) also has the antecedents of ruleIDs 1, 3 and 4 occurring before the HOI5 recorded on 22/3/2010, so its total number of HOI5 association rules is 3 and its maximum confidence, chi-squared and lift are 0.03, 200 and 1.5 respectively. The final instance 4 does not experience any HOI5 association rule antecedent prior to the HOI5 recorded on 5/1/2011, this instance has 0 HOI5 association rules and its maximum confidence, chi-squared and lift are all 0.

As three of the four DOI1-HOI5 instances had a HOI5 association rule with a lift greater than 1, they are considered expected. The absolute risk is 4/25 whereas the confounding-adjusted risk is (4-3)/25=1/25. Considering all the instances, the average instances' maximum confidence and maximum chi-squared are 0.0225 and 150, respectively. Using the real data, these averages will be calculated to give insight into whether the confidence or chi-squared values may be useful to consider for filtering the association rules to use during the refinement. If in this work we find that the chi-squared value seems to be consistently high for signal instances that are unlikely to correspond to ADRs and low otherwise, then this would suggest it is a good measure to use in further research.

In the example we have used a low number of DOI exposures for simplicity, in reality the number of patients prescribed a DOI is likely to be in the thousands.

\subsection{Software}
This research was completed using SQL to access and process the data and the open source analytical software R to do the analysis. The R package `arules'  \cite{arules} was used to do the association rule mining.

\section{Results using THIN}
The drug family of quinolones is used to test the methodology. There are a total of $258397$ first time prescriptions of a quinolone in the database. Following discussions with clinical experts we selected four different health outcomes that had an association with quinolones within the THIN database and applied the refinement methodology to refine each instance. The health outcomes chosen are described in Table \ref{sig} and were chosen to provide a fair range of events. During implementation we discovered that it was not currently computationally efficient to mine association rules with antecedents containing more than three items.

\subsection{Candidiasis (AB2)}
The HOI `Candidiasis' was selected, as it is a known side effect of any antibiotic medication. Out of all the HOIs investigated in this paper the measure of association was lowest for AB2. The Read code AB2 was recorded within 1 and 60 days following the first prescription of a quinolone for 3388 prescriptions. Therefore there were a total of 3388 quinolone-AB2 instances. 

Over 6 million association rules were generated of the form \{antecedent\} $=>$ \{AB2\} with a minimum confidence constraint of 0.01 and a minimum left support constraint of 0.001. For 1293 of the quinlone-AB2 instances, the patient’s basket containing all the medical records prior to the ‘AB2’ record contained one or more AB2-association rule's antecedents. 1293 of the quinlone-AB2 instances' baskets also contained the antecedent of an AB2-association rule with a lift greater than 1. This suggests up to 38\% of the instances may have a non-quinolone cause. The average of the instances' maximum confidences and average of the instances' maximum chi-squared values were 0.31 and 16383 respectively (or 0.82 and 42927 respectively when ignoring the instances whose basket containing all the medical records prior to the ‘AB2’ record did not contained any AB2-association rule antecedent).

The absolute risk value for AB2 during two months after a prescription of a quinolone is $3388/258397=1.2 \times 10^{-2} $ , whereas the confounding-adjusted risk is $(3388-1293)/258397=8.1 \times 10^{-3} $ .

Examples of the rules with the greatest lift are {vaginal discharge, urine pregnancy test, antifungal-drugs}, 

\subsection{Sudden death, cause unknown (R21)}
The HOI `Sudden death, cause unknown' was chosen as it has a temporal bias. Death cannot occur before a patient takes a drug, so using the AB ratio will often result in an association between a drug and Read code corresponding to death. By investigating this Read code, we can see whether it is possible to remove this form of temporal bias via the refinement method.

The Read code R21 was recorded during 1 and 60 days after a first prescription of a quinolone for 73 prescriptions. This corresponds to a total of 73 quinlone-R21 instances. There were 11,219 association rules generated in the form \{antecedent\}$=>$ \{R21\} with a minimum confidence of 0.01 and a minimum left support of 0.001. For 58 of the quinlone-R21 instances, the patient’s basket containing all the medical records prior to the ‘R21’ record contained one or more R21-association rule's antecedents. 58 of the quinlone-R21 instances' baskets also contained the antecedent of an R21-association rule with a lift greater than 1. This suggests up to 79\% of the instances may have a non-quinolone cause. The average instances' maximum confidence and average instances' maximum chi-squared was 0.012 and 564 respectively (or 0.015 and 710 respectively when ignoring the instances whose basket containing all the medical records prior to the ‘R21’ record did not contained any R21-association rule antecedent).

The absolute risk value for R21 during two months after a prescription of a quinolone is $73/258397=2.8 \times 10^{-4} $ , whereas the confounding-adjusted risk value is only $(73-58)/258397=5.8 \times 10^{-5} $.

Examples of the rules with a lift greater than 16 are: (Death, Depression, male) $=>$ Sudden onset of death; (Depression, heart failure, symptoms affect the skin) $=>$ Sudden onset of death and  (Vaccination, heart failure, symptoms affect the skin) $=>$ Sudden onset of death.

The majority of the rules contained death, depression and heart failure. The association with death is expected, but the death recording is likely to occur after the recording of sudden death or on the same day. Therefore, these rules are unlikely to be used during the refinement, as we only look at the medical records that occur before the recording of the sudden death. 

\subsection{Multiple organ failure (C19)}
The HOI `multiple organ failure' was chosen, as it is a severe outcome that occurred 9 times during the month after the first prescription of a quinolone and never in the month before. This HOI is rarely recorded; C19 was recorded into the database for only 262 patients.

Apply association rule mining with a left minimum support of 0.001 and a minimum confidence of 0.01 retuned only one rule of the form \{antecedent\}$=>$\{C19\}. The rule was \{Screening general, septicaemia and tropical corticosteroids\} $=>$\{C19\}. The antecedent corresponding to this association rule was not present in the baskets corresponding to any of the 9 instances of C19 after a quinolone. Therefore, none of the instances was refined, as shown in Table \ref{summ}.

The absolute risk value for C19 during two months after a prescription of a quinolone is $9/258397=3.5 \times 10^{-5} $ , and this is the same as the confounding-adjusted risk.

\subsection{Secondary malignant neoplasm (B572)}
The HOI `Secondary malignant neoplasm' was selected, as it is extremely unlikely to correspond to an acute adverse reaction to any drug. B572 was recorded five times during the month after a first prescription of a quinolone and never in the month before. Therefore is appears to be associated to the quinolones.

There were over 2 million association rules generated of the form \{antecedent\}$=>$ \{B57\} (we only generate rules for the first three elements of the Read code, but if a patient is susceptible to ‘B57’ then they will also be susceptible to ‘B572’) with a minimum confidence constraint of 0.01 and a minimum left support constraint of 0.001. For 5 of the quinlone-B572 instances, the patient’s basket containing all the medical records prior to the ‘B572’ record contained one or more B57-association rule's antecedents. 5 of the quinlone-B572 instances' baskets also contained the antecedent of a B57-association rule with a lift greater than 1. This suggests up to 100\% of the instances may have a non-quinolone cause. The average of the instances' maximum confidences and average of the instances' maximum chi-squared values were 0.045 and 2710 respectively.

The absolute risk value for B572 during two months after a prescription of a quinolone is $5/258397=1.9 \times 10^{-5} $ , whereas the confounding-adjusted risk is 0.

Table \ref{summ} presents a summary of the results for the four Read codes investigated.
\begin{table*} \centering \caption{The results of the novel method for the four quinolone-HOI signals investigate in this paper.} \label{summ}
\begin{tabular}{cccc cc}
HOI & Read code & AB ratio & Instances & Risk & Confounding-adjusted risk \\ \hline \hline
Candidiasis & AB2 &1.27 &3388 &$1.2 \times 10^{-2}$ & $8.1 \times 10^{-3}$ \\
Sudden death, cause unknown & R21& 73& 73& $2.8 \times 10^{-4}$ & $ 5.8 \times 10^{-5}$ \\
Multiple organ failure & C19& 9& 9 & $3.5 \times 10^{-5}$& $3.5 \times 10^{-5}$ \\
Secondary malignant neoplasm & B572& 5& 5& $1.9 \times 10^{-5}$ & 0 \\
\end{tabular}
\end{table*}

\section{Discussion}
The results show that using association rules to refine instances of a signalled DOI-HOI pair may help reduce the number of DOI-HOI pairs that require formal evaluation and may highlight DOI-HOI that are very likely to correspond to ADRs. Interestingly the results show that the refinement method was able to show that the Read code B572 corresponding to secondary malignant neoplasm is an unlikely ADR to the quinolones as the association rules were able to show an alternative cause of the Read code B572 for all the instances. However, the refinement method struggled with the Read code C19, corresponding to multiple organ failure due to the rarity of the Read code resulting in only one association rule with C19 as the consequence. The fact that it may be difficult to generate association rules for rarely recorded Read codes is a limitation of the methodology, however, when a rarely recorded Read code such as C19 occurs fairly often after a DOI, then this is suspicious and probably sufficient reason to formally evaluate it. 

The Read code AB2 corresponding to candidiasis is a known ADR to the quinolones. The Read code was recorded frequently during the two months after a quinolone was prescribed for the first time and is also commonly recorded in the database in general. The high number of medical baskets containing AB2 is probably the reason why there were so many association rules containing AB2 as the consequence. Interestingly, even though there were a large number of association rules, only 1293 instances were identified to have a potential alternative cause. It was suspected that when there were a large number of association rules, then this might lead to the refinement method being overly sensitive and filtering the majority of the instances. This was not the case and further supports the idea that association rules can be used to refine DOI-HOI signals.

The refinement method will not work for patients with a short medical history prior to the DOI prescription, as it requires using items the patient had before the HOI and if they are newly registered, then they are unlikely to have many items recorded. There is little that can be done to solve this issue, as even medical experts would be extremely unlikely to be able to classify the HOI as suspect with no or little medical history for the patient.

In this paper we did not consider the likelihood of the Read code being prescribed based on the age of the patient corresponding to each instance. This may be a useful factor to consider in the future, as some Read codes may be recorded frequently when patients are a certain age and a patient's year of birth is always recorded into the database.

The HOI association rules were generated using a set minimum left support and minimum confidence. It is probably better to tune these parameters based on how common the HOI and DOI are. For rare DOI and HOIs, the minimum values may need to be reduced, whereas they may need to be increased when the HOI and DOI are common. The confidence values of the rules tend to be very small, with the majority of the rules having a confidence less than 5\%. This suggests that confidence is a poor measure of interestingness to apply for this application of association rules. Possible reasons for the confidence being low in this study is that the majority of the HOIs investigated are fairly rare and have multiple causes, so the number of baskets containing an antecedent and HOI will be very small compared to the number containing the antecedent. The lift seems to be more appropriate, as this accounts for the rarity of the HOI.

This study is also limited by restricting the size of the antecedent due to computational issues. Only association rules with antecedents containing three of fewer items were generated. These rules were generated in minutes, resulting in the proposed refinement method being highly efficient when the number of antecedents is limited to 3 or less. By implementing association rule mining using high performance or parallel computing, it may be possible to remove this restriction and still efficiently learn rules with antecedents containing a larger number of items. This may help refine more instances and improve the confounding-adjusted risk value.

Overall, this paper has shown a proof of concept that it may be possible to use association rules to filter DOI-HOI signals that occur due to confounding, but there is still a large amount of research required before the methodology is practical. 

\section{Conclusions}
In this paper we presented a proof of concept for a novel efficient ADR signal refinement method that filters instances of a DOI-HOI signal and does not require knowledge of possible confounders. The recorded history of a patient experiencing the signal is used to filter instances where the medical event can be explained by alternative causes (other than the drug).

The tentative results suggest that the method has the capability to efficiently refine ADR signals but each signal may require specific tuning to determine the optimal support and confidence values to be implemented. Suggestions for future work involve developing methods to efficiently tune the support and confidence parameters and implementing the methodology using distributed computing technology to enable association rules containing large sets of antecedents to be mined. 

Future areas of work could investigate incorporating age to remove DOI-HOI signals caused by age confounding, investigate using different measures of association rule interestingness and develop ways of tuning the method based on how common the HOI/DOI are.

\bibliographystyle{IEEEtran}
\bibliography{refs}

%

\end{document}